\documentclass[12pt]{iopart}
\usepackage{citesort}
\usepackage{graphicx}
\usepackage{amssymb}
\usepackage{xcolor}
  
\begin{document}

\title{Spin polarization in Lateral two-dimensional Heterostructures}

\author{S. Hannan Mouasvi$^1$, and H. Simchi$^2$}

\address{$^1$ Electrical Engineering Department, Islamic Azad University, Central Tehran Branch, Tehran, Iran}
\address{$^2$ Department of Physics, Iran University of Science and Technology, Narmak, Tehran 16844, Iran}
\ead{hannan.mousavi@gmail.com}
\vspace{10pt}
\begin{indented}
\item November 2020
\end{indented}

\begin{abstract}
In this work, we study the spin polarization in the $MoS(Se)_{2}-WS(Se)_{2}$ Transition metal dichalcogenide heterostructures by using the non-equilibrium Green's function (NEGF) method and a three-band tight-binding model near the edges of the first Brillouin zone. Although it has been shown that the structures have no significant spin polarization in a specific range of energy of electrons, by applying a transverse electric field in the sheet of the metal atoms, shedding light on the sample, and under a small bias voltage, a significant spin polarization in the structure could be created. Besides, by applying a suitable bias voltage between leads and applying the electric field a noticeable spin polarization can be found even without shedding the light on the heterostructures.
\end{abstract}

\section{\label{intro}Introduction}
By decreasing the minimum feature size of electronic devices, the charge-based ones encounter some limitations including power dissipation and the manufacturing cost. There are different strategies for overcoming these kinds of limitations. Beyond the Moore strategy, scientists have tried to substitute not only novel materials instead of silicon but also the spin degree of freedom instead of the charge of an electron. After discovering the single atomic layer of graphite, called graphene \cite{1}, due to some graphene boundaries, scientists tried to discover and use the graphene-like two dimensional materials as an alternative of the conventional three dimensional semiconductors for manufacturing electronic devices. Transition metal dichalcogenides (TMDs) with general formula as $MX_{2}$ (M=Mo, W and X=S, Se) are one group of them. TMDs have fascinating and unique properties such as strong spin-orbit coupling (SOC) in the valence band edge, the number of layers dependent bandgap, higher absorption coefficient per unit thickness as compared to conventional semiconductors \cite{2}, long spin lifetime \cite{3,4,5,6}, and strong excitonic effects \cite{7}. Regarding the strong spin-orbit coupling, significant conduction and valence band spin splitting, and  very long electron spin lifetimes, TMDs are one of the main candidate for manufacturing the spin-based devices \cite{8,9,10} and many scientific groups are trying to improve the specifications of TMDs. For instance, Guo et al., have used doping techniques to enhance the spin splitting in monolayer of $WS_{2}$ \cite{3}. It has been shown that the spin relaxation time is enhanced by a small in-plane magnetic field in monolayer of $WSe_{2}$ \cite{11}. Komsa et al., tuned the bandgap and in consequence the spectral region of $MoS_{2}/MoSe_{2}/MoTe_{2}$ compounds by alloying \cite{12}. Based on the assertion of another group, the size and direct-indirect bandgap in the $WX_{2}$ (X=S, Se, and Te) monolayer can be altered by strain engineering technique \cite{13}. Stability enhancement of $MoS_{2}$ by changing its phase from 2H to 1T through Li intercalation is the subject of Ref.\cite{14}. Forming stacked $MX_{2}$ (M=Mo, W and X=S) heterostructures as it has been asserted in Ref.\cite{15} is a way to expedite electron-hole separation for photonic applications. Besides, some references have reported a correlation between the formation of heterostructures and altering the photocatalytic and optical properties of TMDs \cite{7,16}.\\
In the process of fabricating electronic devices, a homojunction is created by doping an impurity in the semiconductor materials whilst a heterojunction, as the basis of heterostructures, is made by using two semiconductors with different energy gaps. Generally, TMD heterostructures are divided into two lateral and vertical groups. At least two layers of different TMD monolayers are needed to be stacked and form a vertical (planar) heterostructure. Similarly, If the layers stitch beside each other, a lateral (in-plane) heterostructure can be formed. In this regard, various groups have been reported practical achievement of sharp and defect-free lateral heterostructures via different methods such as lateral epitaxial and thermal CVD processes \cite{6,14,17}. This is while improving the interfacial properties of the heterostructures by means of various methods is still in progress \cite{6}. A wide range of TMD based applications has been reported so far, e.g. in-plane transistors, diodes, p-n photodiodes, and complementary metal-oxide-semiconductor (CMOS) inverters \cite{6}. Furthermore, optoelectronic applications of van der Waals (vdW) and lateral heterostructures are increasing due to the rapid interlayer charge transfer \cite{18} and type engineering ability \cite{6}. Heo et al., have reported a rotation-based interlayer photoexcitation in $MoS_{2}/WS_{2}$ monolayer stacks due to the in-direct to direct transitions \cite{19}. Likewise, the enhancement of photoexcited charge carriers separation and collection through photovoltaic structure under lateral and vdW heterostructures was carried out by Atwater et al \cite{2}.\\
A fine review on spintronic in two dimensional materials and their heterostructures can be found in Ref.\cite{20}. It has been shown that the heterostructures can significantly improve the SOC \cite{20}. Some examples are graphene with a WS2 substrate, TIs/graphene vdW heterostructure, and MoS2 on the graphene channel \cite{20}. Zhang et al., have reported high-spin polarization at room temperature in 2D layers by reducing its carrier lifetime via the construction of vdW heterostructures. A near unity degree of spin polarization has been observed in PbI2 layers with the formation of type-I and type-II band aligned vdW heterostructures with monolayer of $WS_{2}$ and $WSe_{2}$ \cite{21}. Zhou et al., have studied the vdW graphene/hafnene heterostructure with different stacking configurations and shown that the distinct electronic distribution and spin-polarized characteristic of the vdW heterostructures come from the intensity of orbital overlap induced by diverse stacking configurations \cite{22}. Deyan et al., have reviewed and discussed some of the distinctive effects observed in ferromagnetic junctions with prominent 2D crystals such as graphene, hexagonal boron nitride, and transition metal dichalcogenides and showed how spin interface phenomena at such junctions affects the observed magnetoresistance in devices \cite{23}. By referring to another available references like Refs.\cite{24,25,26}, one can study more about different applications of TMD heterostructures as well as prefect spin and valley polarization in some TMD-based monolayers.\\
Using the NEGF method, spin polarization in lateral $MoX_{2}-WX_{2}$ (X=S, Se) heterostructures has been studied by our team in this paper. Besides, the domination of W (Mo)-atom orbitals on the negative (positive) energy range of electrons for both types of spin degree of freedom, and the main role of $d_{{z}^{2}}$ orbital of both W- and Mo-atoms in the quantum conductance of the device under the equilibrium condition (no bias voltage between leads) have been revealed by our studies. Under equilibrium condition and without applying a transverse electric field and shedding light on the structure, no significant spin polarization was seen when the energy of an electron was less than -0.1 eV. On the other hand, under the non-equilibrium condition plus applying the field and shedding light on the sample, in a specific range of electron energy, due to the breaking of the time-reversal symmetry, a significant spin polarization was seen in both $MoSe_{2}-WSe_{2}$ and $MoS_{2}-WS_{2}$ lateral heterostructures. In addition, it has been shown, the $d_{{z}^{2}}$ orbital of edge atoms plays the main role in the quantum conductance of the device under the cited condition. In consequence, we show that a significant spin polarization can be seen in some TMD heterostructures under special conditions and give a general positive answer to the probable future application of spin degree of freedom in the TMD heterostructures \cite{24,25,26}, as it has been demonstrated that a heterostructure can improve the SOC \cite{27}. It is worth noting, up to our best knowledge, the spin polarization in lateral TMD heterostructures haven't been studied theoretically and experimentally. In consequence, we can not compare our results with others. \\
The structure of this article is as follows. The material parameters and calculation methods are provided in sections \ref{matpar} and \ref{calc}, respectively. Section \ref{rslt} includes the results and discussion, and in section \ref{sum} one can read the summary.\\

\section{\label{matpar}Material parameters}
\begin{figure}[t]
 \centering
 \includegraphics[scale=1]{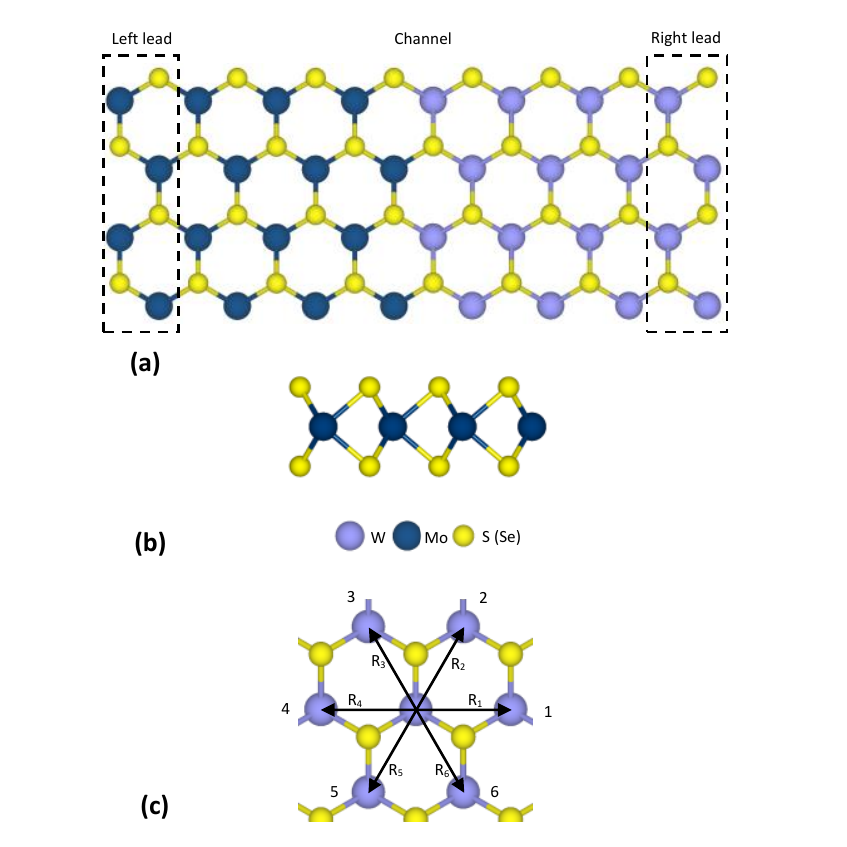}
 \caption{(Color online) Schematic of $MoS(Se)_{2}-WS(Se)_{2}$ TMD heterostructure $(a)$ top view, $(b)$ left lead side view, $(c)$ next nearest neighborhood atoms respect to each central metal atom. $\protect\overrightarrow{R}_{i}$(i=1,…,6) are the displacement vectors. The lattice constant and the length of each $\protect\overrightarrow{R}_{i}$ vector are equal to 3.19 $\mathring{A}$ (3.34 $\mathring{A}$) in $MoS(Se)_{2}-WS(Se)_{2}$ heterostructure.}
 \label{fig1}
 \end{figure}

The honeycomb configuration of the $MoS(Se)_{2}-WS(Se)_{2}$ TMD heterostructures which are illustrated in Fig.\ref{fig1} are composed of three sections including the left lead, right lead, and channel. In our calculations, the channel area which is divided into two $MoS(Se)_{2}$ and $WS(Se)_{2}$ semi-infinite nanoribbons at the left and right side appropriately, consists of 72 metal atoms. Note that the mismatch between the two semi-infinite nanoribbons is negligible \cite{28}. Fig.\ref{fig1} (c) presents the next nearest neighborhood atoms respect to a central metal atom. The magnitude of each displacement vector, $\overrightarrow{R}_{i}$(i=1,…,6), in that figure equals to 3.19 $\mathring{A}$ (3.34 $\mathring{A}$) for $MoS(Se)_{2}-WS(Se)_{2}$ heterostructure. It must be emphasized, in the numerical calculations, we only consider the metal atoms with a triangular lattice structure, and in consequence, electrons only hop between the metal atoms in six different directions.

\section{\label{calc}Calculation method}
Exploiting a three-band Hamiltonian model \cite{29,30,31,32,33,34} and the NEGF method, besides the spin polarization studies in lateral two- dimensional $MoS(Se)_{2}-WS(Se)_{2}$ TMD heterostructures, it has been shown, at the two K and K’ corners of first Brillouin zone (BZ), the Bloch states in monolayer $MX_{2}$ (M=Mo, W and X=S, Se) mostly consist of $d_{{z}^{2}}$ orbitals for the conduction band minimum (CBM), and $d_{xy}$ and $d_{{x}^{2}-{y}^{2}}$ for the valence band maximum (VBM) \cite{5,35,36,37,38,39}. Additional information including the hopping matrix between metal atoms in each displacement direction and on-site matrix of each metal atom is provided in appendix \ref{TB}.
In order to calculate the spin polarization and spin current which are defined as Eq.\ref{eq:SP} and Eq.\ref{eq:IEV} respectively, one should evaluate the transmission probability, T +(-), of spin up (down) electrons through the NEGF method \cite{40,41,42,43}.\\

\begin{eqnarray}
\fl{SP=\frac{T^{+}-T^{-}}{T^{+}+T^{-}}} \label{eq:SP}
\end{eqnarray}
\begin{eqnarray}
\fl{I}^{+(-)}(E,V)=\frac{q}{\pi{}\hslash{}}\int{T}^{+(-)}(E,V)[\mathit{f}_{1}(E,V)-\mathit{f}_{2}(E,V)]dE \label{eq:IEV}
\end{eqnarray}

\begin{figure}[b]
 \centering
 \includegraphics[scale=1]{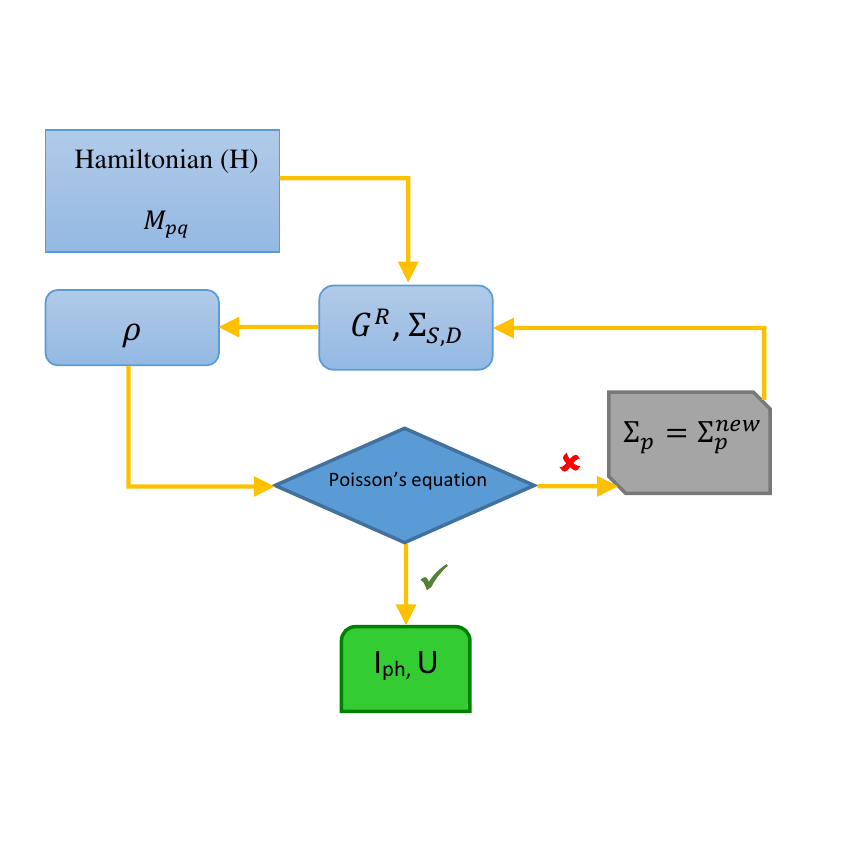}
 \caption{(Color online) The self-consistent method for calculating the effect of shedding light on the heterostructure.}
 \label{fig2}
 \end{figure}

In Eq.\ref{eq:IEV}, $\hslash{}$, q, E, V, and $\mathit{f}_{1,2}$ refer to the reduced Planck constant, elementary charge of electron, energy, potential, and Fermi distribution function of leads, respectively. Near the equilibrium condition, by shedding the light on the channel, some electrons are excited to higher energy levels and their energy increases by $\hslash{}\omega{}$ where $\omega{}$ is the frequency of the incident light. After a while, they come back to lower energy levels due to losing the same energy. In consequence, we are calculating the effect of the incident light on the quantum conductance of a two levels system \cite{40} in which the energy of its electrons is equal to $E\pm\hslash{}\omega{}$. Using Fermi’s golden rule \cite{44}, one can calculate the electron-photon coupling matrix and consequently the self-energy of incident light to implement the influence of it on the system. This could be done by adding the mentioned self-energy to the Green’s function of the electrons similar to the effect of leads and their self energies (Appendix \ref{Pint}). Althuogh the detailed information about incident light self-energy calculations exists in Appendix B, a brief explanation is provided in the continuation of this paragraph. In the first step, the electron and hole correlation functions and the initial value of carrier’s density are calculated by using the Green’s function of the channel and self-energies of leads without shedding the light (zero self-energy of the photon). By means of the coupling matrix and correlation functions, one can calculate the initial value of photon self-energy which is used for evaluating the new amount of the Green’s function of the channel and the self-energy of the photon. Using Poisson’s equation, the loop is repeated until the convergence is achieved i.e., it is a self-consistent method (Appendix \ref{Pint}) in accordance with Fig.\ref{fig2}.\\

Since the transverse electric field is applied in the plane of metal atoms, its effect is added to the diagonal elements of the Hamiltonian matrix as a potential term. The mentioned term is calculated by multiplying the strength of the field to the y-coordinate of each atom while the lower (upper) edge of the nanoribbon is considered as the reference line. In addition, the bias voltage between leads changes the Fermi distribution function at left (right) lead equal to $\mathit{f}(E+V_{DS}/2) (\mathit{f}(E-V_{DS}/2))$ where $V_{DS}$ is bias voltage. The difference between these two distributions i.e., $[\mathit{f}(E+V_{DS}/2)-\mathit{f}(E-V_{DS}/2)]$ should be considered in calculating the transmission probability \cite{40}. It should be noted, both the bias voltage and shedding light, place the system in the non-equilibrium condition. Thus, the self-consistent method should be used for calculating the transmission probability and spin polarization \cite{40}. It is important to note that we used the optimal values for energy of photons, intensity of light, and the intensity of $E_{T}.y$ which are 0.12 eV, $20 nW/\mu m^{2}$, and 0.2 eV, respectively in our mentioned numerical calculations. In other words, we found that, by altering the mentioned parameters, the spin polarization could be changed. Therefore, first, we calculated the optimum values for them aimed at getting the maximum spin polarization, and then the optimal values were exploited in our calculations.\\

\section{\label{rslt}Results and discussion}
\begin{figure}[b]
 \centering
 \includegraphics[scale=1]{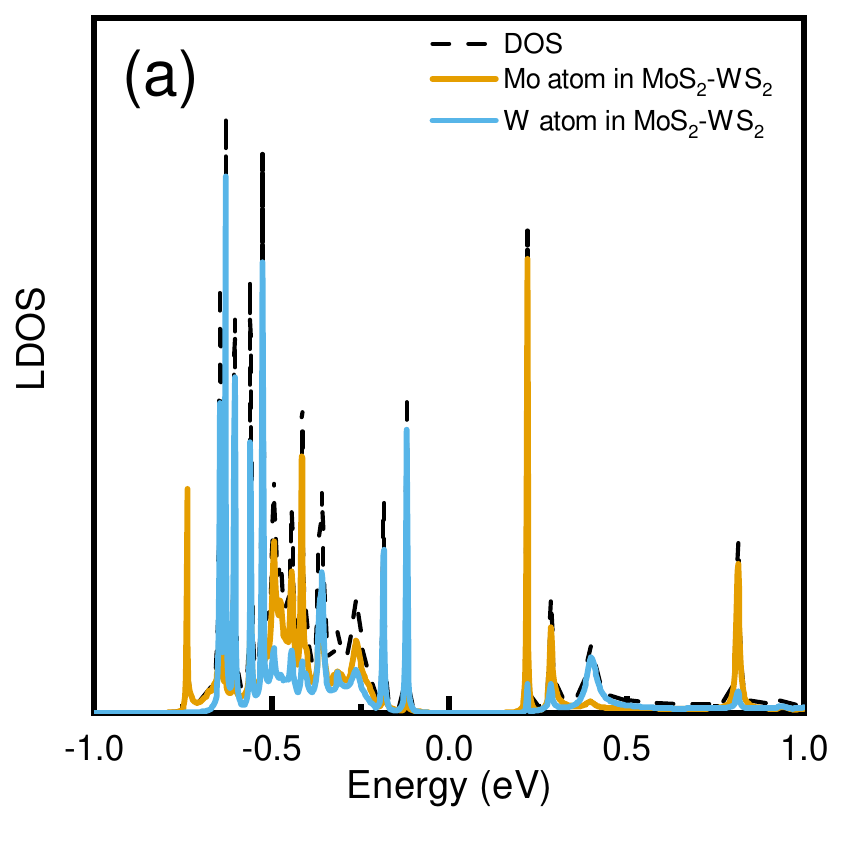}
  \includegraphics[scale=1]{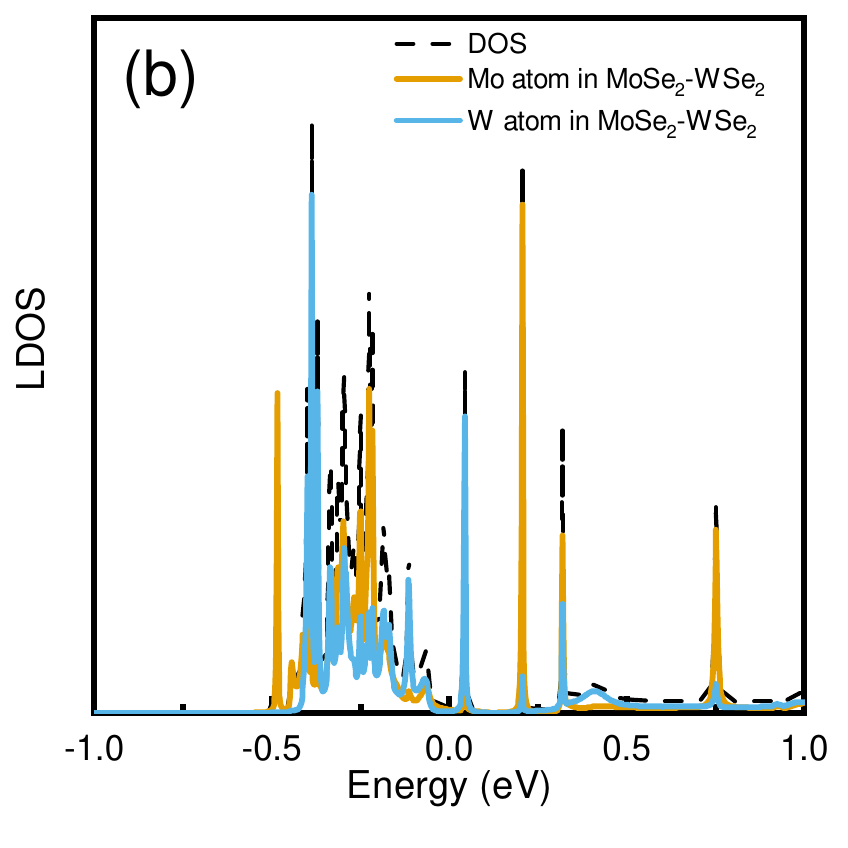}
 \caption{(Color online) Spin up localized density of states $(a)$ $MoS_{2}-WS_{2}$ and $(b)$ $MoSe_{2}-WSe_{2}$, under equilibrium condition, i.e., without light, electric field, and bias voltage.}
 \label{fig3}
 \end{figure}

Fig.\ref{fig3} (a) and (b) show the spin up localized density of states (LDOS) for $MoS_{2}-WS_{2}$ and $MoSe_{2}-WSe_{2}$ TMD heterostructures under equilibrium condition, respectively. As these figures show, the LDOS of Mo (W)-atom dominates on the positive (negative)  range of energy of electrons. Although by using density functional theory (DFT) Kang et al., have shown that the highest occupied molecular orbital (HOMO) belongs to W-atom and the lowest unoccupied molecular orbital (LUMO) belongs to Mo-atom \cite{45}, because of considering the molecular orbitals of metal atoms, our results differ from their results. Density of state (DOS) which is the summation of LDOSs, is shown in the dashed line in Fig.\ref{fig3} (a) and (b) for $MoS_{2}-WS_{2}$ and $MoSe_{2}-WSe_{2}$ heterostructures accordingly. Based on the figures, one can say that the DOS is zero for -0.1 to 0.2 eV electron energy range in $MoS_{2}-WS_{2}$ and -0.02 to 0.02 eV in $MoSe_{2}-WSe_{2}$ heterostructures. In consequence, it is expected that a transmission gap will be seen in the curve of the quantum conductance.\\

 \begin{figure}[b]
 \centering
 \includegraphics[scale=1]{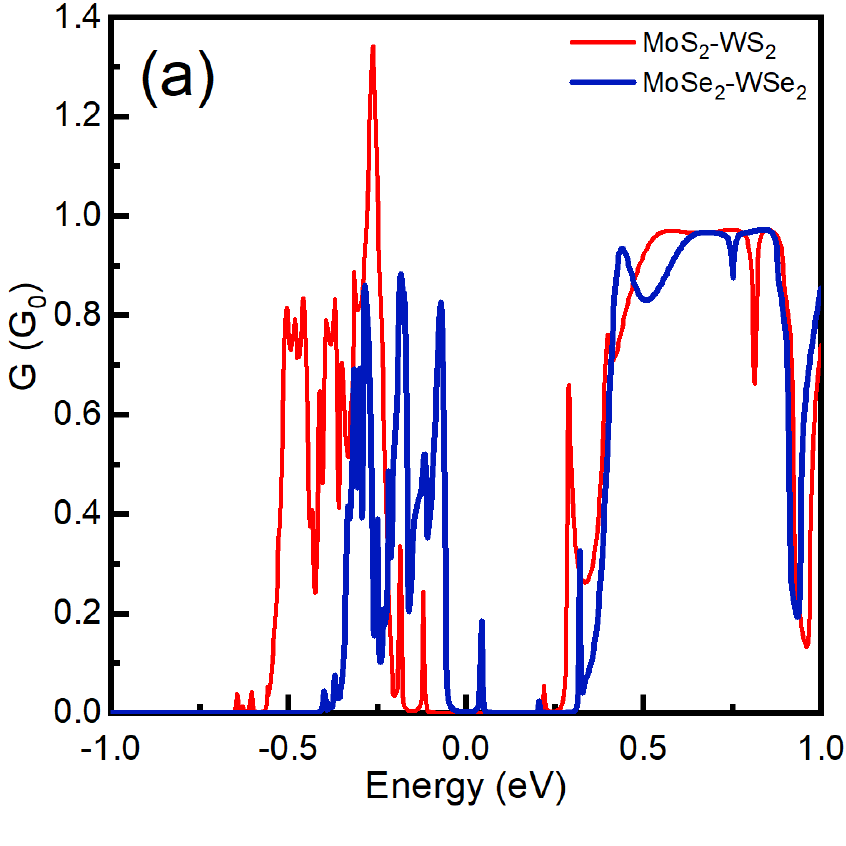}
  \includegraphics[scale=1]{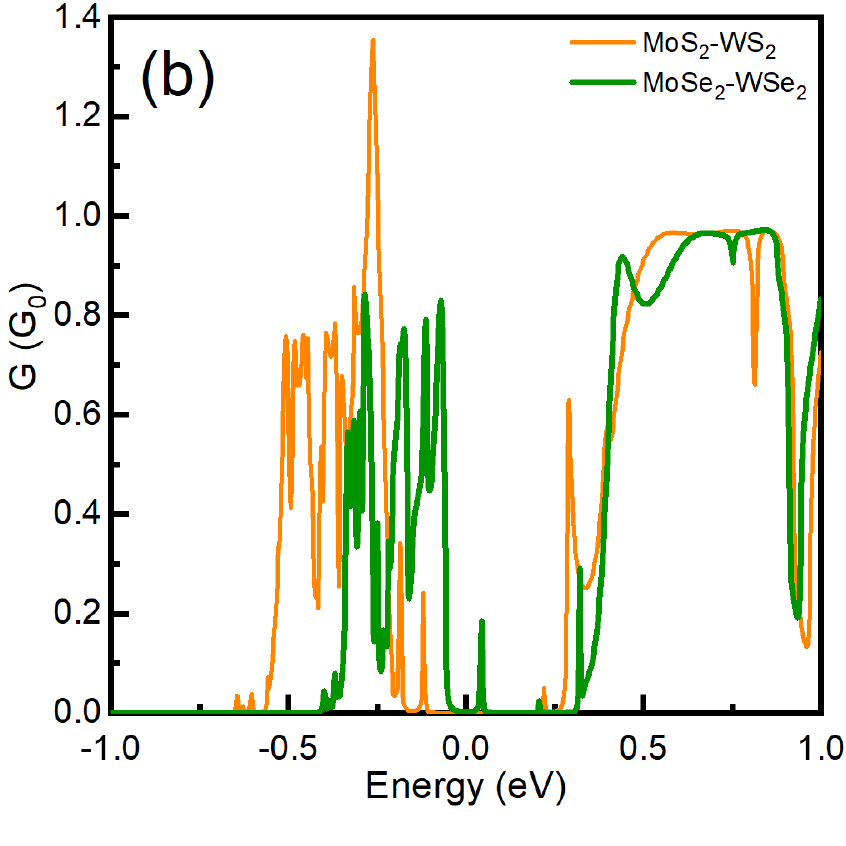}
 \caption{(Color online) Quantum conductance of $MoS(Se)_{2}-WS(Se)_{2}$ for $(a)$ spin up and $(b)$ spin down electrons, uner equilibrium condition.}
 \label{fig4}
 \end{figure}
 
 \begin{figure}[h]
 \centering
 \includegraphics[scale=1]{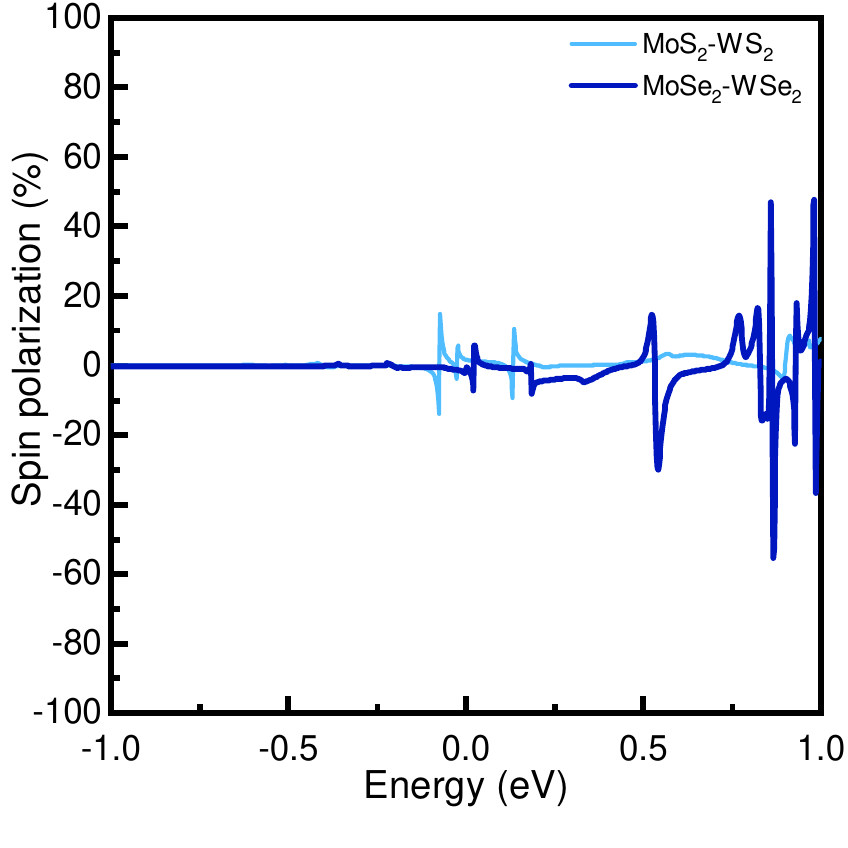}
 \caption{(Color online) Spin polarization for $MoS(Se)_{2}-WS(Se)_{2}$ without light, electric field, and bias voltage.}
 \label{fig5}
 \end{figure}
 
\begin{figure}[t]
 \centering
 \includegraphics[scale=1]{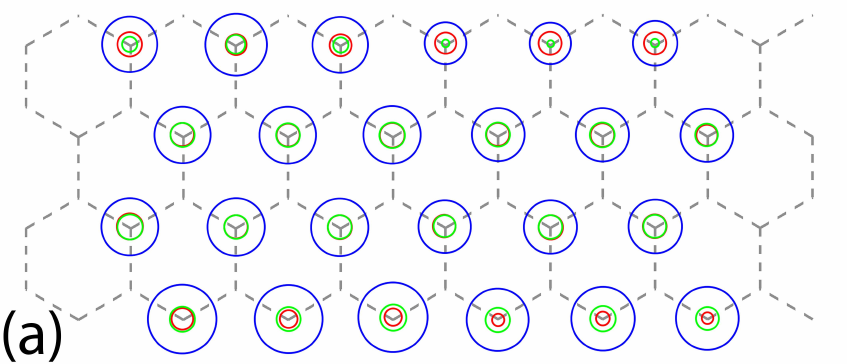}
 \includegraphics[scale=1]{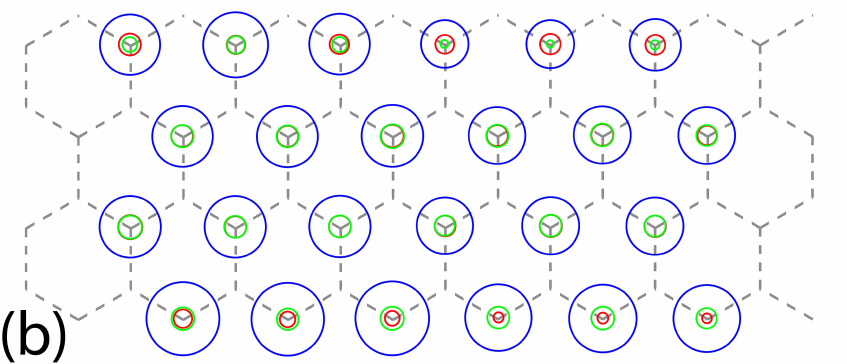}
 \caption{(Color online) The magnitude of LDOSs for spin up electrons $(a)$ $MoS_{2}-WS_{2}$ and $(b)$ $MoSe_{2}-WSe_{2}$ under equilibrium condition. The $d_{{z}^{2}}$ orbital is shown in blue color, $d_{xy}$ orbital in red color and $d_{{x}^{2}-{y}^{2}}$ in green color.}
 \label{fig6}
 \end{figure}

\begin{figure}[t]
 \centering
 \includegraphics[scale=1]{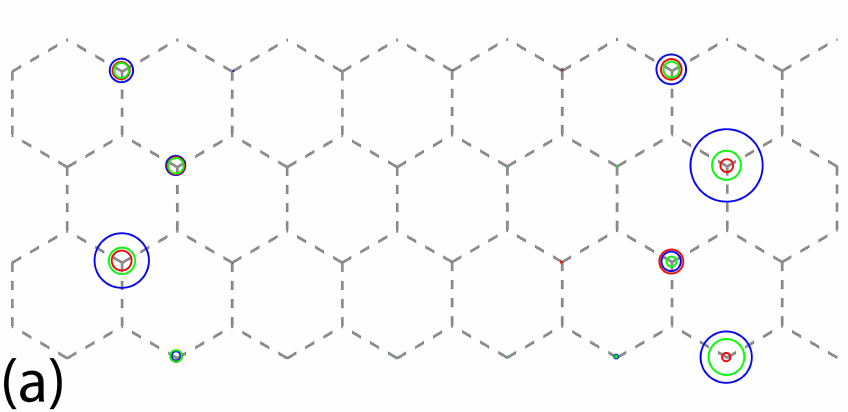}
 \includegraphics[scale=1]{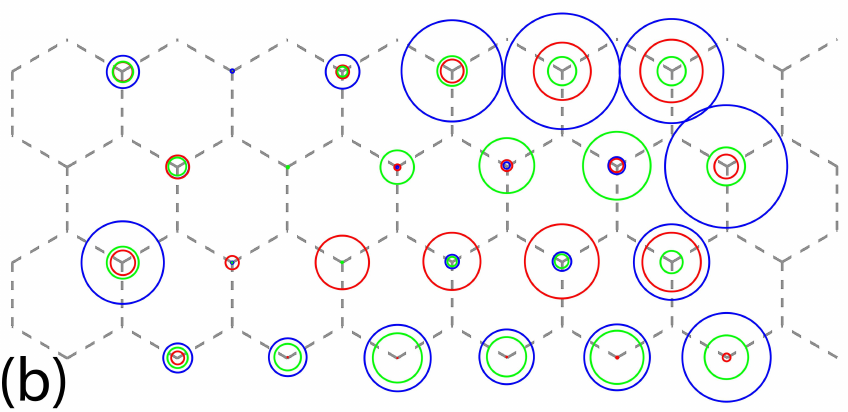}
 \caption{(Color online) The magnitude of LDOSs after shedding the light for spin up electrons $(a)$ $MoS_{2}-WS_{2}$ and $(b)$ $MoSe_{2}-WSe_{2}$ under equilibrium condition. The $d_{{z}^{2}}$ orbital is shown in blue color, $d_{xy}$ orbital in red color and $d_{{x}^{2}-{y}^{2}}$ in green color.}
 \label{fig7}
 \end{figure}

To calculate the quantum conductance of $MoS_{2}-WS_{2}$ and $MoSe_{2}-WSe_{2}$ TMD heterostructures for both spins in accordance with Fig.\ref{fig4} (a) and (b) respectively, we used the DOS of the system. As the mentioned figures show, when the energy of an electron, $E_{e}\geqslant -0.25 eV$, a small difference is seen between the conductance of spin-up electrons and spin-down ones. Correspondingly, a transmission gap equals to 0.3 (0.04) eV is seen in $MoS_{2}-WS_{2}$ $(MoSe_{2}-WSe_{2})$ heterostructure. This implied that both nanoribbons are narrow bandgap semiconductors. Based on the curve of spin polarization under equilibrium condition as Fig.\ref{fig5}, for $E_{e}\geqslant -0.25$eV, spin polarization is not zero, generally. For better understanding the role of atomic orbitals on the quantum conductance, we show the magnitude of LDOSs for spin-up electrons as circles on the lattice position of each atom (Fig.\ref{fig6}). As the figure shows, the $d_{{z}^{2}}$ orbital of all atoms has the main role in the quantum conductance not only in W-side but also in Mo-side. According to Fig.\ref{fig6} (a) and (b), the appeared difference in the quantum conductance curves can be attributed to the different roles and the radius of blue circles on the position of W and Mo atoms which are placed at the upper edge of a nanoribbon. Fig.\ref{fig7} shows the LDOS after shedding the light on the sample. Comparing Fig.\ref{fig7} with Fig.\ref{fig6}, one can conclude that the LDOS decreases due to the occupation of states by excited electrons, probably. It should be noted, in order to have a good view of LDOS changes in the presence of light as well as make the mentioned comparison possible, the circles' diameters are illustrated about four times larger than the calculated amounts.\\

 \begin{figure}[t]
 \centering
 \includegraphics[scale=1]{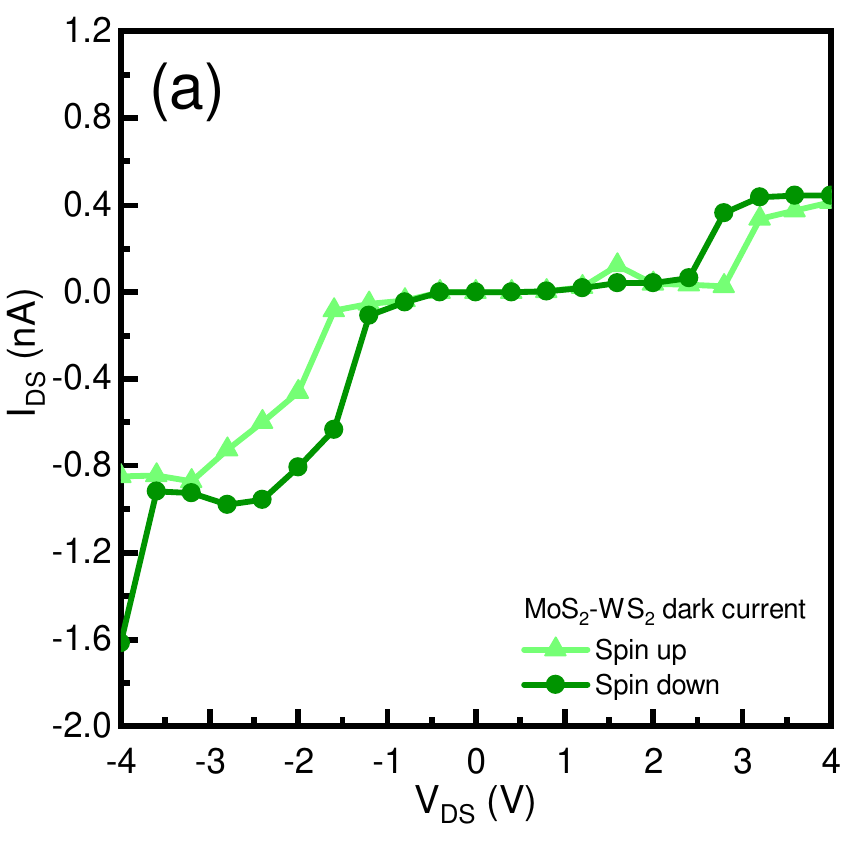}
 \includegraphics[scale=1]{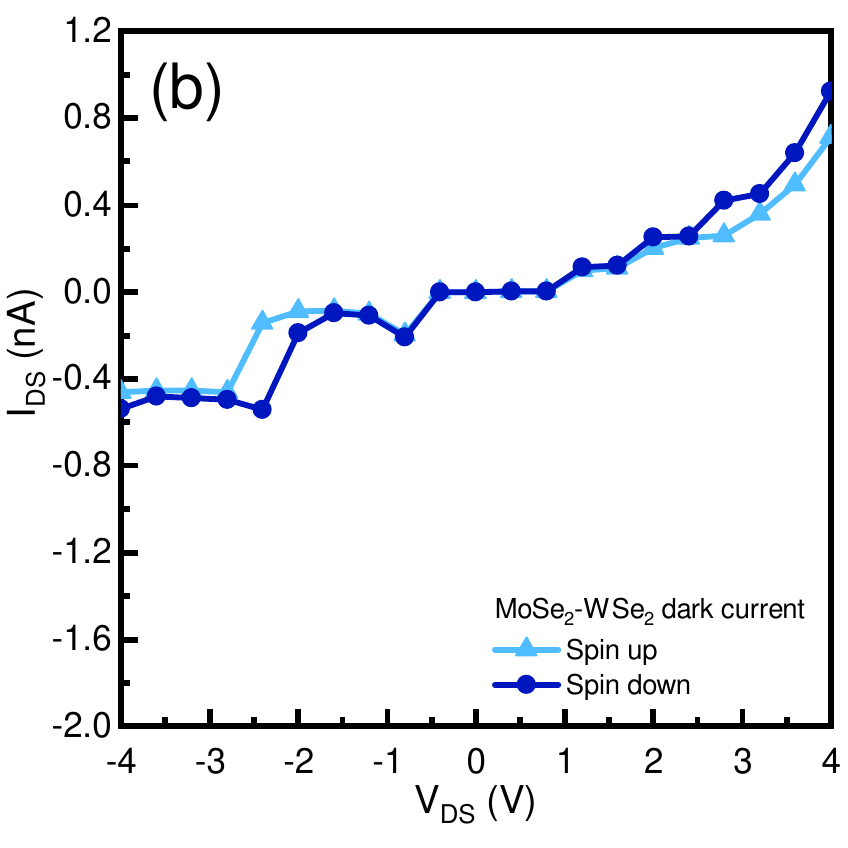}
 \caption{(Color online) Spin current versus bias voltage $(a)$ $MoS_{2}-WS_{2}$ and $(b)$ $MoSe_{2}-WSe_{2}$ nanoribbon. Here, the $E_{T}=0.2 eV$.}
 \label{fig8}
 \end{figure}

Creating spin polarization in a structure could be done by implementing an external factor such as electric field \cite{46}, magnetic field \cite{25,47,48}, exchange field \cite{49}, or bias voltage \cite{50} and consequently breaking time-reversal symmetry (TRS). With that mean, first, we apply a bias voltage at a range from -4 to 4 volts and a transverse electric field equal to 0.2 eV and calculate the spin-dependent current. As the spin-dependent current versus bias voltage curve in Fig.\ref{fig8} shows, for some values of bias voltage, the spin-dependent current is not zero and therefore, the spin polarization has a non-zero value. A negligible value of the spin polarization for both $MoS_{2}-WS_{2}$ and $MoSe_{2}-WSe_{2}$ heterostructures in energies from 0.41eV to 0.45eV has been achieved (see Fig.\ref{fig5}). Moreover, as presented in Fig.\ref{fig8}, for $-0.1V \leqslant V_{DS} \leqslant 0.1V$, the spin-up current is equal to spin down current approximately. Now, a question can be asked. Is it possible under these conditions i.e., $0.41 eV \leqslant E_{e} \leqslant 0.45 eV$ and $-0.1V \leqslant V_{DS} \leqslant 0.1V$, in which the non-zero spin polarization changes to a significant spin polarization value under shedding the light on the channel and applying a transverse electric field ($E_{T}$)?\\

\begin{figure}[b]
 \centering
 \includegraphics[scale=1]{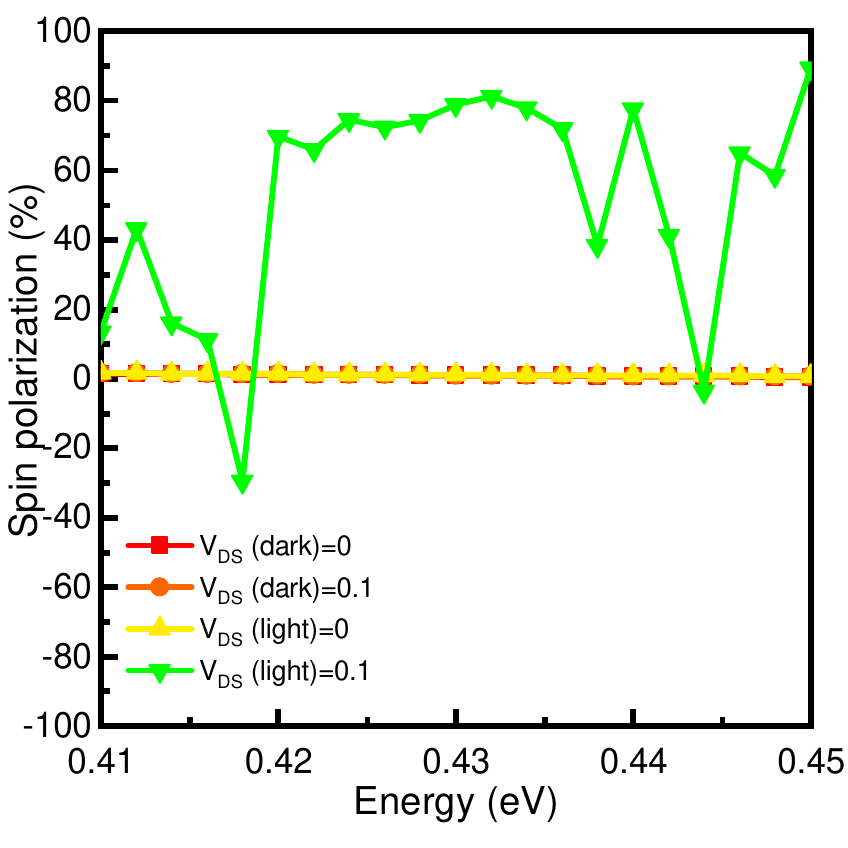}
 \caption{(Color online) Spin polarization in $MoSe_{2}-WSe_{2}$ under light shedding, 0.2 eV electric field, and various bias voltage. The intensity of light is $20 nW/\mu m^{2}$ and its energy is 0.12 eV. d (l) means dark (light).}
 \label{fig9}
 \end{figure}

\begin{figure}[t]
 \centering
 \includegraphics[scale=1]{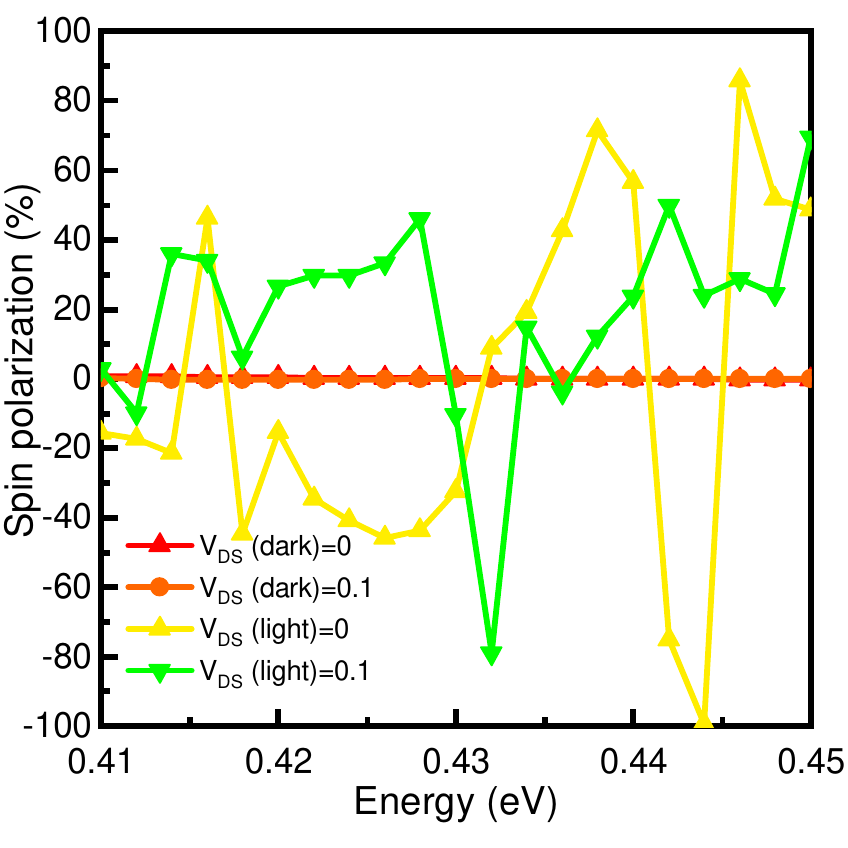}
 \caption{(Color online) Spin polarization for $MoS_{2}-WS_{2}$ under light, 0.2eV electric field, and various bias voltage. The intensity of light is $20 nW/\mu m^{2}$ and its energy is 0.12 eV. d (l) means dark (light).}
 \label{fig10}
 \end{figure}

As illustrated in Fig.\ref{fig9}, which is the curve of spin polarization versus the energy of electrons, when $V_{DS}=0.1 V$, $E_{T}.y=0.2 eV$, and the channel is under shedding the light with $20 nW/\mu m^{2}$ intensity and 0.12 eV   energy, a significant and almost constant spin polarization is seen for $0.42 eV \leqslant E_{e} \leqslant 0.44 eV$. It means that the answer to the above question is positive when the $MoSe_{2}-WSe_{2}$ heterostructure is considered. Whilst, in accordance with Fig.\ref{fig10}, with shedding light on the channel of $MoS_{2}-WS_{2}$ heterostructure, a non-zero spin polarization can be seen.\\

To clarify the subject with regard to the influence of spin-orbit coupling in VBM and CBM degeneracy, spin-valley dependency, and optical selection rule in $W(M)X_{2}$ monolayers \cite{51}, one can say that the light excites the electrons from non-degenerate VBM to non-degenerate CBM based on the spin-valley dependency and optical selection rule. In the excitation process by the light the difference between the density of spin up and spin down electrons changes and in consequence, it is expected that a non-zero spin polarization is seen. As Fig.\ref{fig9} shows, when the energy of electron, $E_{e}$, is approximately in a range from 0.42 to 0.44 eV, the spin polarization in a $MoSe_{2}-WSe_{2}$ heterostructure has a nearly constant positive value (0.6 to 0.8) due to the domination of spin-up carriers while $E_{T}.y=0.2 eV$, $V_{DS}=0.1 V$, and light with the previous specifications exists on the channel. In accordance with the previous one, the $MoS_{2}-WS_{2}$ TMD heterostructure has no spin polarization under the same condition while the channel is dark. In other words, by applying $E_{T}$ and $V_{DS}$ simultaneously and without shedding the light on the channel, the difference between the densities of spin up and spin down electrons is negligible, whereas with and without $V_{DS}$ the spin polarization is created while the light and $E_{T}$ are present. Similarly, the light excites the electrons based on the spin-valley dependency and optical selection rule such that the difference between the density of spin-up electrons and spin-down electrons changes. It should be noted that the perfect spin polarization is seen with $E_{T}$, shedding the light, and without bias for $E_{e}=0.444 eV$ (Fig.\ref{fig10}).\\

\section{\label{sum}Summary}
Exploiting the NEGF method and three-band Hamiltonian model, we studied the spin polarization in $MoS_{2}-WS_{2}$ and $MoSe_{2}-WSe_{2}$ lateral heterostructures. Under the equilibrium conditions, It was shown that non-significant spin polarization happens when the energy of an electron ($E_{e}$) is from 0.42 to 0.44 eV, in both heterostructures, whereas by applying a transverse electric field, bias voltage, and shedding light on the $MoSe_{2}-WSe_{2}$ structure, a significant spin polarization was seen for $0.42 eV \leqslant E_{e} \leqslant 0.44 eV$. Conversely, it was shown that a significant spin polarization happens only by shedding the light on the bias-less heterostructure while a transverse electric field exists.\\
 
\appendix
\section{\label{TB}Tight-binding details}
The Bloch states in monolayer $MX_{2}$ (M=Mo, W and X=S, Se) mostly consist of $d_{{z}^{2}}$ orbitals for conduction band minimum (CBM), and $d_{xy}$ and $d_{{x}^{2}-{y}^{2}}$ for the valence band maximum (VBM) \cite{5,35,36,37,38,39}. Therefore, by considering the base functions 
$|\Phi_{1}>=d_{{z}^{2}}$, $|\Phi_{2}>=d_{xy}$, $|\Phi_{3}>=d_{{x}^{2}-{y}^{2}}$ the next nearest tight- binding Hamiltonian \cite{35} can be written as equation \ref{eq:Mat1}.\\
\begin{eqnarray}
\fl H^{NN}(K)=
\left( \begin{array}{ccc}
h_{0} & h_{1} & h_{2} \\
h_{1}^{*} & h_{11} & h_{12} \\
h_{2}^{*} & h_{12}^{*} & h_{22} \\
\end{array} \right)
\label{eq:Mat1}
\end{eqnarray}
Each element of $H^{NN}$ is a $3 \times 3$ matrix that represents the hopping integrals \cite{52}. These hopping integrals change in the different displacement directions which are shown in Fig.\ref{fig1} (c).  The hopping matrices are as \ref{eq:Mat2} to \ref{eq:Mat4} relations \cite{52}.\\
\begin{eqnarray}
\fl h(R_{1})=
\left( \begin{array}{ccc}
t_{0} & t_{1} & t_{2} \\
-t_{1} & t_{11} & t_{12} \\
t_{2} & -t_{12}^{*} & t_{22} \\
\end{array} \right)
=h(R_{4})^{\dagger}
\label{eq:Mat2}
\end{eqnarray}
\begin{eqnarray}
\fl h(R_{6})=
\left( \begin{array}{ccc}
t_{0} & \frac{1}{2}t_{1}-\frac{\sqrt{3}}{2}t_{2} & -\frac{\sqrt{3}}{2}t_{1}-\frac{1}{2}t_{2} \\
-\frac{1}{2}t_{1}-\frac{\sqrt{3}}{2}t_{2} & \frac{1}{4}t_{11}+\frac{3}{4}t_{22} & \frac{\sqrt{3}}{4}t_{11}-t_{12}-\frac{\sqrt{3}}{4}t_{22} \\
\frac{\sqrt{3}}{2}t_{1}-\frac{1}{2}t_{2} & -\frac{\sqrt{3}}{4}t_{11}+t_{12}+\frac{\sqrt{3}}{4}t_{22} & \frac{3}{4}t_{11}+\frac{1}{4}t_{22} \\
\end{array} \right)
=h(R_{3})^{\dagger}
\label{eq:Mat3}
\end{eqnarray}
\begin{eqnarray}
\fl h(R_{2})=
\left( \begin{array}{ccc}
t_{0} & \frac{1}{2}t_{1}+\frac{\sqrt{3}}{2}t_{2} & \frac{\sqrt{3}}{2}t_{1}-\frac{1}{2}t_{2} \\
-\frac{1}{2}t_{1}+\frac{\sqrt{3}}{2}t_{2} & \frac{1}{4}t_{11}+\frac{3}{4}t_{22} & -\frac{\sqrt{3}}{4}t_{11}-t_{12}+\frac{\sqrt{3}}{4}t_{22} \\
-\frac{\sqrt{3}}{2}t_{1}-\frac{1}{2}t_{2} & \frac{\sqrt{3}}{4}t_{11}+t_{12}-\frac{\sqrt{3}}{4}t_{22} & \frac{3}{4}t_{11}+\frac{1}{4}t_{22} \\
\end{array} \right)
=h(R_{5})^{\dagger}
\label{eq:Mat4}
\end{eqnarray}
Therefore, the three-band TB Hamiltonian will be rewritten as relation \ref{eq:Mat5}: \\
\begin{eqnarray}
\fl H=
\left( \begin{array}{ccc}
h_{D} & \cdots & h(R_{n}) \\
\vdots & \ddots & \vdots \\
h(R_{n})^{\dagger} & \cdots & h_{D} \\
\end{array} \right)
\label{eq:Mat5}
\end{eqnarray}
where, \\
\begin{eqnarray}
\fl h_{D}(up)=
\left( \begin{array}{ccc}
\epsilon_{1} & 0 & 0 \\
0 & \epsilon_{2} & i\lambda \\
0 & -i\lambda & \epsilon_{2} \\
\end{array} \right)
\label{eq:Mat6}
\end{eqnarray}
\begin{eqnarray}
\fl h_{D}(down)=
\left( \begin{array}{ccc}
\epsilon_{1} & 0 & 0 \\
0 & \epsilon_{2} & -i\lambda \\
0 & i\lambda & \epsilon_{2} \\
\end{array} \right)
\label{eq:Mat7}
\end{eqnarray}
The constants of the above equations are defined in Table \ref{tbl1} \cite{34}.
\begin{table}[h]
\caption{GGA based three-band TB model parameters of monolayer MX2 in the unit of eV \cite{34}}
\centering 
\begin{tabular}{llllllllll}
\hline\hline
   & \multicolumn{1}{c}{$\epsilon_{1}$} & \multicolumn{1}{c}{$\epsilon_{2}$} & \multicolumn{1}{c}{$t_{0}$} & \multicolumn{1}{c}{$t_{1}$} & \multicolumn{1}{c}{$t_{2}$} & \multicolumn{1}{c}{$t_{11}$} & \multicolumn{1}{c}{$t_{12}$} & \multicolumn{1}{c}{$t_{22}$} & \multicolumn{1}{c}{$\lambda$}    \\
\hline
$MoS_{2}$ & 1.046 & 2.104 & -0.184 & 0.401 & 0.507 & 0.218 & 0.338 & 0.057  & 0.073 \\
$MoSe_{2}$ & 0.919 & 2.065 & -0.188 & 0.317 & 0.456 & 0.211 & 0.290 & 0.130  & 0.091 \\
$WS_{2}$ & 1.130 & 2.275 & -0.206 & 0.567 & 0.536 & 0.286 & 0.384 & -0.061 & 0.211 \\
$WSe_{2}$ & 0.943 & 2.179 & -0.207 & 0.457 & 0.486 & 0.263 & 0.329 & 0.034  & 0.228\\
\hline
\end{tabular}
\label{tbl1}
\end{table}\\

\section{\label{Pint}Light interaction}
The retarded Green’s function $G^{R}$ is defined as:\\
\begin{eqnarray}
\fl G^{R}=[(E+i\eta)\mathbf{I}-H_{c}-\Sigma_{l1}-\Sigma_{l2}-\Sigma_{photon}]^{-1} \label{eq:GRN}
\end{eqnarray}
Where $\eta$ equals $1.5 \times 10^{-3}$, $\mathbf{I}$ is the unit matrix, $\Sigma_{l1,2}$ are self-energy of leads, and $\Sigma_{photon}$ is photon self-energy. The effect of leads on channel appears as shifting the energy levels and broadening the energy level.
It is assumed, due to the energy of photons, $\hslash{}\omega{}$, the energy of electrons changes as E+($\hslash{}\omega{}$) and E-($\hslash{}\omega{}$) under shedding the light on the channel, and in consequence, similar to the leads the effect of photons can be added to the Hamiltonian of the system as a self-energy. In order to evaluate the mentioned self-energy, in the first instance, we assumed that the self-energy of a photon in that equation is zero. This allows us to calculate the initial value of $\rho$ by:
\begin{eqnarray}
\fl \rho=dE [diag(\mathit{f}_{L}*G^{R}*\Gamma_{l1}*G^{A})+diag(\mathit{f}_{R}*G^{R}*\Gamma_{l2}*G^{A})] \label{eq:RHO}
\end{eqnarray}
Where, $\mathit{f}_{L,R}$ and $\Gamma_{l1,2}$ are Fermi distribution function and broadening matrix of electrodes, respectively. Using equation \ref{eq:Sph} one can calculate the preliminary value of photon self-energy.\\
\begin{eqnarray}
\fl \Sigma_{photon}=\frac{i[\Sigma_{photon}^{in}+\Sigma_{photon}^{out}]}{-2} \label{eq:Sph}
\end{eqnarray}
Where $\Sigma_{photon}^{in}$ and $\Sigma_{photon}^{out}$ are defined \cite{42} as:
\begin{eqnarray}
\fl \Sigma_{photon}^{in}=\sum_{pq}\quad M_{lp}*M_{qm}*[NG_{pq}^{n}(E-\hslash\omega)+(N+1)G_{pq}^{n}(E+\hslash\omega)] \label{eq:Sph1}
\end{eqnarray}
\begin{eqnarray}
\fl \Sigma_{photon}^{out}=\sum_{pq}\quad M_{lp}*M_{qm}*[NG_{pq}^{p}(E+\hslash\omega)+(N+1)G_{pq}^{p}(E-\hslash\omega)] \label{eq:Sph2}
\end{eqnarray}
Where $M_{qm}$, N, $G^{n(p)}$, and $\hslash\omega$ are electron-photon coupling matrix, number of photons, electron (hole) correlation function, and energy of photons, respectively. With the aim of  calculating the proper value of photon self-energy, we have to calculate the potential of the system through Poisson’s equation as:
\begin{eqnarray}
\fl U=(-\rho)/P \label{eq:POT}
\end{eqnarray}
Where P is the discretized Poisson operator (appendix \ref{PsnEq}). Finally, we compare the new potential value with the initial one. If their difference is bigger than a specific arbitrary value, we guess the new self-energy of photon and substitute it in equation \ref{eq:GRN}. This self- consistent loop continues until the mentioned potential difference satisfies the mentioned requirements.\\
\section{\label{PsnEq}Poisson’s equation}
\begin{figure}[t]
\centering
\includegraphics[scale=1]{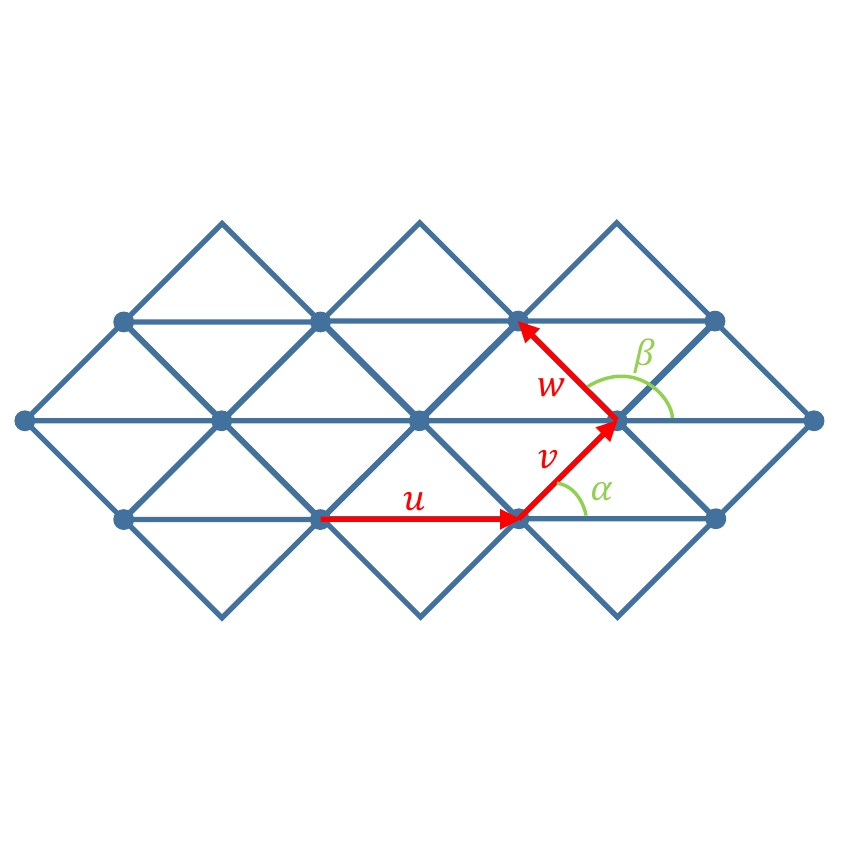}
\caption{(Color online) The triangular lattice and definition of new vectors u, v, and w.}
\label{figc-1}
\end{figure}

Poisson’s equation in Cartesian geometry has a form like equation \ref{eq:PSN}.\\
\begin{eqnarray}
\fl \frac{\partial^{2}u}{\partial x^{2}}+\frac{\partial^{2}u}{\partial y^{2}}+\mathit{f}=0 \label{eq:PSN}
\end{eqnarray}
With the aim of changing the Cartesian geometry to Triangular one, three new vectors called u, v, and w vectors as shown in Fig.\ref{figc-1} are defined. It should be noted, the lattice of metal atoms is triangular and in consequence, the angles $\alpha$ and $\beta$ are equal to 60 and 120 degrees, respectively. Therefore, the Eq.\ref{eq:PSN} can be written as \cite{53}:\\
\begin{eqnarray}
\fl \frac{\partial^{2}u}{\partial x^{2}}+\frac{\partial^{2}u}{\partial y^{2}}=\frac{2}{3}(\frac{\partial^{2}u}{\partial u^{2}}+\frac{\partial^{2}u}{\partial v^{2}}+\frac{\partial^{2}u}{\partial w^{2}}) \label{eq:PSN1}
\end{eqnarray}

Using the finite difference method, the Eq.\ref{eq:PSN1} can be discretized and equation \ref{eq:FD} will be found \cite{54}:\\
\begin{eqnarray}
\fl -6\Phi_{i,j}+\Phi_{i,j-2}+\Phi_{i,j+2}+\Phi_{i+1,j+1}+\Phi_{i-1,j+1}+\Phi_{i+1,j-1}+\Phi_{i-1,j-1}=Q_{i,j}+O(h^{6})\label{eq:FD}
\end{eqnarray}
\\
\\
\\
\\

\bibliographystyle{iopart-num}

\bibliography{Refs}

\end{document}